\documentclass[twocolumn,showpacs,aps,prb]{revtex4}
\usepackage{epsfig}

\begin{document}
\author{M. Shuker\footnotemark \footnotetext{Email
address: shuker@physics.technion.ac.il}, A. Ben-kish, A. Fisher,
and A. Ron} \affiliation{Department of Physics, Technion - Israel
Inst. of Tech., Haifa 32000, Israel}
\title{Titanium plasma source for capillary discharge extreme ultraviolet lasers}
\begin{abstract}
A technique to generate jets of pure Titanium plasma is presented. A Ti wire is exploded in an Alumina capillary sealed with 1atm. of air inside. The generated plasma emerges from the capillary (to a high-vacuum environment) by ripping a thin Ti foil that seals one of the capillary ends. The generated plasma jets have a velocity  of up to
$4.5\pm0.5mm/\mu s$, an electron temperature of $1.5\pm0.5eV$ and
a ion density of $2.7\pm1\times10^{17}/cc$. The plasma source was designed for a capillary discharge extreme ultraviolet laser experiment, but might also be useful to other application such as a target for Z-pinch experiments.
\end{abstract}
\maketitle
\begin{figure}
    \epsfig{file=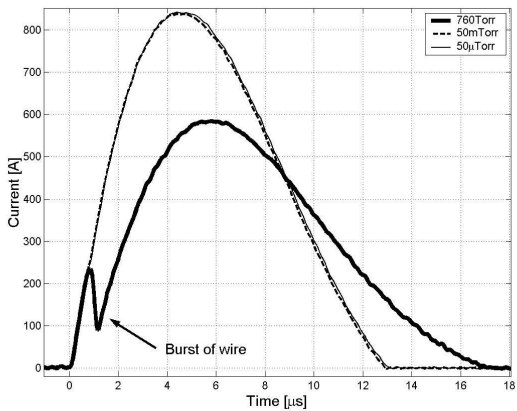}
    \caption{Current signals of exploding wire in atmospheric pressure (thick line), in 50mTorr (dashed line) and in 50mTorr (thin line). A signature of the wire explosion is evident only in atmospheric pressure.}
    \label{FigureCurrents}
\end{figure}

\newpage
Among several techniques explored for achieving short wavelength
lasing \cite{Elton1990}, the capillary discharge has the advantage
of being relatively simple, efficient and compact
\cite{Rocca1994},\cite{Rocca1999}. This technique is based on an
electrical discharge preformed on low density gaseous targets in a
dielectric capillary. Recently, a capillary discharge extreme ultraviolet (XUV) laser was realized
in the $3P\rightarrow3S$ transition of Ne-like Ar ions at
$46.9nm$. This laser was initially demonstrated by Rocca and his
colleagues \cite{Rocca1994}, and several years later by other
groups \cite{BenKish2001},\cite{Niimi2001},\cite{Tomassetti2002}.
One of the challenges in scaling the capillary discharge laser to
shorter wavelength is the formation of the initial plasma target.
The target should be a capillary uniformly filled with low density
($N_i\cong10^{16} par/cc$), pre-ionized gas or vapor. This can be
easily achieved with gases (or liquids with high vapor pressure
\cite{Frati2000}), however targets from materials that are solid
at room temperature are more difficult to implement. Furthermore, high purity and uniformity targets are required as the laser is sensitive to instabilities.\\
Several possible techniques to create suitable plasma targets were
previously explored. Ablative capillaries were used as the main
capillary \cite{Rocca1993} or as a secondary capillary shooting a
jet of plasma into the main capillary \cite{Tomasel1997}. The
ablation can take place during the main current pulse or by adding
a pre-pulse. In the case the required material is a conductor, an
insulating compound must be used and an impure plasma is created
(for example, a pressed $TiH_2$ capillary was used to create a Ti
plasma target \cite{Rocca1993}). Another possible solution is
generating the plasma by ablating the capillary electrodes
\cite{Kukhlevsky1997}, \cite{Rahman2003}. This technique is
suitable for conductors, but is not efficient and might result in
ablating part of the capillary walls as well.
\\In this letter we address the problem of creating a metallic
plasma source. For our experiments we used Ti whose Ne-like ion
lase at $32.6nm$. We adopt the concept of the secondary
capillary \cite{Tomasel1997}, and suggest using an exploding wire
in the secondary capillary to produce the plasma (instead of
ablating the walls or the electrodes).The exploding wire has the
advantages of being very efficient and producing high purity Ti
plasma (its major drawback is being a single-shot device). We note that the plasma jet must be created in a vacuum environment
(atmospheric density is a few order of magnitude higher than the
required density of the active atoms). \\In order to determine the
desired properties of the Ti plasma target (i.e. the
properties just prior to the onset of the main current pulse), the
dynamical process of the capillary discharge must be considered.
We used a one dimensional magneto-hydrodynamical (MHD) code
\cite{Nemirovsky1999}, to model the capillary
discharge experiment and to infer the required properties of the $Ti$ plasma
target. In all the calculations below we used the parameters of
our pulsed power system \cite{BenKish2001}, that can produce peak
currents of $40-130KA$ with half-cycle time of $100-120ns$. In a
series of calculations we found suitable parameters of the plasma
target that will result in a Ne-like Ti XUV laser. The required plasma target is a capillary filled with pre-ionized ($T_e=T_i=0.1-0.5eV$) Ti plasma, with ion density of $N_i=1-4\times10^{16}/cc$. Therefore, the plasma jet from the secondary capillary must have a substantially higher density ($>10^{17}/cc$) to create a suitable plasma target (it is very simple to reduce the density in the main capillary by a suitable flow geometry \cite{Shuker1998}).\\
The experimental system for generating the Ti plasma target
consisted of a low energy discharge system with a $10\mu F$
capacitor that was typically charged to $2kV$. A low inductance
semi-rigid coax cable was used to transfer the electrical energy
to the secondary capillary. The total inductance of the electrical
system was $900nH$ and the typical half-cycle duration was $10\mu
s$. The peak currents in the system were in the range $0.5-2kA$
(depending on the charging voltage and the capillary setup). The
capillary consisted of a plastic structure with an alumina insert.
The inner diameter of the capillary was 1mm, and its length was
$10-30mm$. The electrodes of the capillary were made of Ti to
minimize the contamination of the Ti plasma. One of the electrodes
had a concentric hole with 1mm diameter (through which the plasma
jet will emerge). The wire in the capillary was a high purity Ti
wire of $125\mu m$ diameter. The experiments were constructed
inside a vacuum chamber - and preformed in various environments
ranging from atmospheric pressure to $10^{-6}Torr$. Electrical measurements during the wire explosion were performed
with a commercial Rogovsky coil and a Tektronix 1:1000 high
voltage probe. A fast visible-light camera was used to study the
flow of the plasma jet from the secondary capillary and its
propagation towards the main capillary. The plasma properties
(composition, electronic density and temperature) were inferred
from visible light spectroscopy measurements using a
spectrometer coupled to a gated CCD camera. \\ The exploding wire technique is a well known method to generate high-temperature plasmas. However, there is a major difficulty in realizing this technique in vacuum environment (as required in our case).
Figure \ref{FigureCurrents} depicts the current signals measured in three experiments performed in our system at different ambient pressure.

\begin{figure}
    \epsfig{file=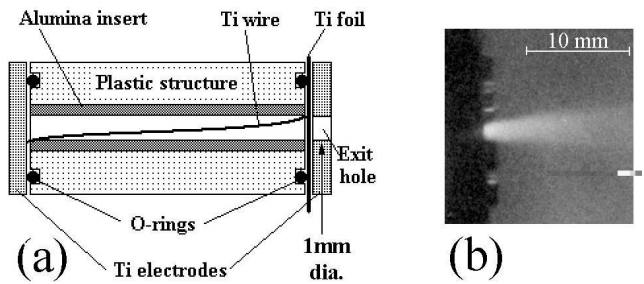}
    \caption{(a) An illustration of the sealed capillary structure. The exit hole is sealed using a thin Ti foil that rips after the wire explosion. (b) Typical fast photography image of a plasma jet emerging from the sealed capillary device (in vacuum).}
    \label{FigureSealedCapillary}
\end{figure}

\begin{figure}
    \epsfig{file=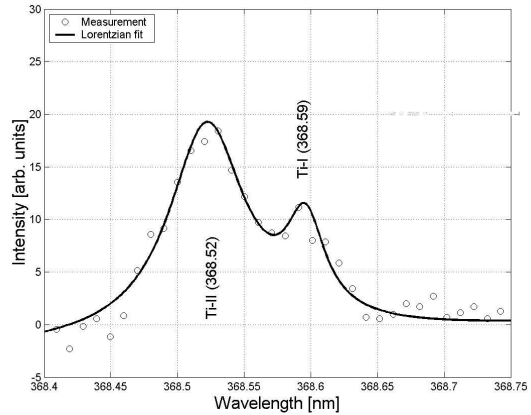}
    \caption{Time integrated, space resolved spectrum of the Ti plasma jet emerging from the sealed capillary. The plasma electron density is inferred from Stark broadening of the Ti-II 368.52nm spectral line.}
    \label{FigureSpectroscopy}
\end{figure}

In atmospheric pressure a clear signature of the wire explosion is
evident (the current sharply drops and increases again shortly
afterwards). However, at ambient pressures $50mTorr,50\mu Torr$
the current shows no evidence of the wire explosion. Fast
photography experiments confirmed that the wire did not explode
(in fact it stayed connected to the electrodes and intact during
the entire electrical discharge), and consequently no plasma was
formed. The reason for this phenomena in vacuum environment is
that when the electrical current starts, it ablates small amount
of material from the wire's outer surface (either material from
the wire bulk or, more likely, material adsorbed to the surface).
Therefore, a low density plasma ($10^{14}-10^{16} /cc$) is formed
around the wire. This low density plasma has lower resistivity
than the wire itself, so a substantial part of the current flows
through the surrounding plasma rather than through the wire. The
reduction in the current that flows through the wire prevents it
from exploding (unless the current pulse is very intense). On the
other hand, when the experiment is performed in atmospheric
pressure the ablated material flows into the high density
($\sim10^{19}/cc$) air surrounding the wire. Hence, low density
plasma cannot be formed and the current continues to flow through
the wire leading to its explosion.
\\ In order to enable the formation of a
plasma jet in vacuum environment a change in the setup of the experiment was introduced.
The inner volume of the capillary was sealed with atmospheric
pressure inside, enabling the explosion of the wire. By
properly selecting the inner diameter of the capillary and the
diameter of the wire the plasma impurity caused by the atmospheric air was kept lower than $1\%$.
The seal of the capillary towards the exit hole
was realized using a thin Ti foil ($12\mu m$). A schematic drawing
of the sealed capillary is presented in figure
\ref{FigureSealedCapillary}.a.
The foil  is designed to stand a pressure difference of more than
$1$ atmosphere without ripping. However, After the wire explodes
the high pressure formed in the capillary rips the foil, and a
plasma jet flows out of the capillary. This capillary was inserted
to vacuum environment (up to $10^{-6}Torr$) and used to eject a
jet of high purity Ti plasma. In all the experiments reported
below we used a $1mm$ inner diameter, $30mm$ long Alumina
capillary with a $125\mu m$ diameter Ti wire. The charging voltage
of the $10\mu F$ capacitor was $2kV$. In all these experiments the
electrical signals showed that the wire exploded properly, and a
jet of Ti plasma emerged from the capillary. The properties of the
Ti plasma jet emerging from the capillary were further
investigated by means of fast photography and visible light
spectroscopy.\\ Fast photography images with exposure time of
$200ns$ were used to study the jet propagation. A typical fast
photography image is displayed in figure
\ref{FigureSealedCapillary}.b, showing the plasma jet formation
outside the sealed capillary nozzle. The velocity of the plasma
jet tip in vacuum was $4.5\pm0.5mm/\mu s$ (compared to
$0.6\pm0.1mm/\mu s$ in similar experiments performed at
atmospheric ambient pressure). Very directional jets with opening
half-angle of $<15^{\circ}$ were produced. The plasma properties
of the jet (electronic temperature and density) were measured
using visible light spectrometery \cite{Griem1964}. The
fluorescence light from the jet was collected using an optical
fiber with an appropriate collimator. The light was collected from
jet at a distance of $1.5\pm1mm$ from the exit hole. The spectrum
was time integrated during the first $80\mu s$ after the plasma
flow started (the long integration time was used in order to
achieve good signal-to-noise ratio). Spectra taken at shorter
integration durations showed similar results, since the flow of
the plasma was steady at these times. We have performed
measurements in several spectral regimes to observe different
lines of Ti and its singly ionized ion. In all the measurements
the majority of the lines were identified as Ti lines, which is an
indication to the purity of the plasma. For a plasma in LTE the
intensity ratio between two spectral line (of the same ion specie)
is given by \cite{Bekefi1976}:
\begin{equation} \label{eqn_intensity_ratio}
\frac{I_1}{I_2}=\frac{f_{ul}(1)g_u(1)}{f_{ul}(2)g_u(2)}(\frac{\lambda_2}{\lambda_1})^3exp[-\frac{E_u(1)-E_u(2)}{k_BT_e}]
\end{equation}
Where the indexes 1,2 stand for the two spectral lines, $I$ is the
measured intensity, $f_{ul}$ is the
absorption oscillator strength, $g_u$ is the statistical weight of
the upper level, $\lambda$ is the wavelength of the transition,
$E_u$ is the energy of the upper level of the transition and $k_B$
is the Boltzman constant. The plasma jet is at LTE since its
electronic density is higher than the critical density (in our conditions $N_{cr}\cong10^{14}/cc$). By comparing the intensities of the
spectral lines $439.4nm$ and $441.7nm$ of natural Ti, and using
eq. \ref{eqn_intensity_ratio} we calculated an electronic
temperature of $1.5\pm0.5eV$.\\ The electronic density was
determined by measuring the broadening of the $368.5nm$ spectral
line of singly ionized Ti. Several mechanisms contribute to the
line-width of a spectral line, including natural broadening,
Doppler broadening, Stark broadening as well as the spectrometer resolution. In the
experimental setup we used, the dominant broadening mechanism for
the $368.5nm$ line was the Strak broadening (electron impact). Therefore the plasmas
electronic density could be determined by measuring the FWHM of
the $368.5nm$ line. In figure \ref{FigureSpectroscopy} the
measured spectrum in the vicinity of the $368.5nm$ line is
depicted.

The spectral line at $368.6nm$ (from natural Ti) also
appears in the spectrum, therefore a double peak curve (Voigt profile) was fitted to both lines (displayed in figure
\ref{FigureSpectroscopy}). From this measurement the electronic
density of the plasma was found to be
$N_e=(4\pm1.5)\times10^{17}elec/cc$ (using electron impact
parameter for the $368.5nm$ line given in \cite{Bartolic1994}).
Similar results were achieved from the broadening of another
spectral line. The mean ionization of Ti plasma at an electronic
temperature of $1.5eV$ is $\overline{Z}=1.5$, therefore the
density of Ti ions in the plasma jet is about
$N_i=(2.7\pm1)\times10^{17}ion/cc$. The measured density at the
exit hole of the capillary is an order of magnitude higher than
the required density of the plasma target. By properly selecting
the geometry of the secondary and main capillaries, a suitable
plasma target can be achieved.\\
In conclusion, a simple technique to produce pure Ti plasma
in vacuum environment was demonstrated. The technique is based on
exploding a $125\mu m$ Ti wire inside a sealed capillary containing air. A thin Ti foil, which seals one of the capillary ends, rips after the wire explosion and
allows the plasma to flow out of the capillary. The Ti plasma jet
formed in the experimental conditions used here had an axial
velocity of $4.5\pm0.5mm/\mu s$, an opening half-angle of
$<15^{\circ}$. A time integrated spectroscopy
measurement performed $1.5\pm1mm$ from the exit hole of the
capillary showed that the plasma had an electron temperature of
$T_e=1.5\pm0.5eV$ and an ion density of
$N_i=(2.7\pm1)\times10^{17}ion/cc$. By properly
selecting the geometry, this plasma jet can be used to create a plasma target suitable for a Ne-like Ti capillary discharge XUV laser
\cite{Shuker2000}. This technique for forming pure metallic plasma
in vacuum might be useful for other applications such as a target
for dense Z-pinch experiments.
\\This work was partially supported by the Fund for
Encouragement of Research in the Technion. We acknowledge the helpful discussions with Ron Nemirovsky and the
technical assistance of Yoav Erlich and Ofer Bokovza.

\end{document}